\documentclass[prb,preprint,showpacs,preprintnumbers,amsmath,amssymb]{revtex4}

\usepackage{graphicx}

\usepackage{dcolumn}

\usepackage{bm}

\newcommand{\comment}[1]{}
\newcommand{\gma}{$\rm{Ga_{1-x}Mn_{x}As}$}

\newcommand{\tc}{$T_{\rm{C}}$}
\newcommand{\tgma}{$t_{\rm{GMA}}$}
\newcommand{\tma}{$t_{\rm{MA}}$}
\newcommand{\tspacer}{$t_{\rm{spacer}}$}
\newcommand{\he}{$H_{\rm{E}}$}
\newcommand{\hc}{$H_{\rm{c}}$}

\begin{document}
\title{Interlayer and interfacial exchange coupling in ferromagnetic metal/semiconductor heterostructures}
\author{M.\ J.\ Wilson,$^ 1$ M.\ Zhu,$^ 1$ R.\ C.\ Myers,$^ 2$ D.\ D. \ Awschalom,$^ 2$ P. \ Schiffer,$^ 1$ and N.\ Samarth$^ 1$}\email{nsamarth@psu.edu}
\affiliation{$^ 1$Dept. of Physics, The Pennsylvania State University, University Park PA 16802\\
$^ 2$Dept. of Physics, University of California, Santa Barbara CA 93106}

\begin{abstract}

We describe a systematic study of the exchange coupling between a magnetically hard metallic ferromagnet (MnAs) and a magnetically soft ferromagnetic semiconductor (\gma) in bilayer and trilayer heterostructures. An exchange spring model of MnAs/\gma~bilayers accounts for the variation of the exchange bias field with layer thickness and composition. We also present evidence for hole-mediated interlayer exchange coupling in MnAs/p-GaAs/\gma~trilayers and study the dependence of the exchange bias field on the thickness of the spacer layer. 

\end{abstract}
\maketitle

\section{Introduction}
The systematic study of interlayer and interfacial exchange coupling in ferromagnetic (FM) metal multilayers has led to important advances in condensed matter physics, in addition to guiding rapid progress in magnetic storage technologies.\cite{Stiles:1999fm,Tsymbal:2001:SSP} In a similar manner, fundamental inquiry into the exchange coupling in FM semiconductor heterostructures could have an important influence on the development of semiconductor spintronics, where we can envision an additional level of opto-electronic control over the underlying exchange interaction.\cite{Awschalom:2007vq} Interlayer and/or interfacial exchange coupling in FM semiconductor heterostructures has indeed been unequivocally observed in a variety of experiments, including neutron scattering measurements of \gma/GaAs multilayers,\cite{Kirby:2007kf,Chung:2008hv} magnetometry and magneto-resistance measurements of MnO/\gma~bilayers \cite{Eid:2004rx,Ge:2007:PRB} and MnAs/\gma~bilayers,\cite{Zhu:2007kt} and x-ray magnetic circular dichroism studies of Fe/\gma~heterostructures.\cite{Maccherozzi:2008kf} However, despite longstanding theoretical interest in this topic,\cite{Jungwirth:1999ee} there is a dearth of systematic experimental data that examines the dependence of this coupling on relevant parameters; the availability of such data is critical for developing a deeper theoretical understanding of interfacial/interlayer exchange coupling in FM semiconductor heterostructures.

Here, we present comprehensive studies of the interlayer and interfacial exchange coupling between a magnetically soft FM semiconductor (\gma) and a magnetically hard FM metal (MnAs). The juxtaposition of these materials is particularly convenient because the interlayer and interfacial exchange coupling are readily measured using standard magnetometry techniques,\cite{Zhu:2007kt} rather than requiring more elaborate methods such as neutron scattering or x-ray magnetic circular dichroism. We systematically map out the variation of the exchange coupling as a function of many sample parameters in both bilayer and trilayer geometries. Our group has previously examined exchange coupling in bilayers as a
function of \gma~ thickness (\tgma), showing that the exchange field (\he) varies inversely with \tgma.\cite{Zhu:2007kt}  In the present work, we first confirm this result in an additional set of samples and then
study a wide range of previously unexplored parameters such as MnAs thickness, magnetization and spacer thickness in trilayer systems, all of which give important insights into the physics of this system. Our results are consistent with the formation of an exchange spring in bilayers due to FM interfacial coupling with MnAs.\cite{Zhu:2008ck} Additionally, in trilayers, we find evidence for FM hole-mediated exchange coupling that decays exponentially with spacer layer thickness, persisting over a length scale of $\sim 5$ nm. We do not see any evidence for antiferromagnetic (AFM) exchange coupling over the entire space of parameters examined. Finally, we find an enhancement of the Curie temperature (\tc) of \gma~layers by the overgrowth of MnAs. Although superficially resembling a ``proximity effect'' wherein the \tc~of a weak ferromagnet might be enhanced by interfacial exchange coupling with a strong ferromagnet,\cite{Maccherozzi:2008kf} control measurements show that in our samples this is an extrinsic effect stemming from the unintentional annealing and gettering of Mn interstitial defects.

\section{Experimental Details}
This paper focuses on four different series of samples.  Series A consists of three bilayer samples with 12 nm ``type-A'' MnAs on top of a \gma~layer with $x \sim 6\%$ and with \gma~layer thicknesses of \tgma$ = 30, 50, 80 $ nm. Series B consists of three MnAs/\gma~bilayers where the composition of the \gma~is varied ($0.05 \leq x \leq 0.16$) while keeping the \gma~and MnAs thicknesses constant at 30 nm and 8 nm, respectively. Series C consists of several MnAs/\gma~bilayer samples ($x \sim 6\%$, \tgma$= 30$nm) in which the MnAs layer is purposely varied in thickness (1 nm $\lesssim$ \tma $\lesssim$ 4 nm)  across the wafer by exposing the static substrate to a spatially inhomogeneous Mn flux. Series D consists of MnAs/p-GaAs/\gma~trilayers (with $x \sim 6\%$) in which the MnAs and \gma~layer thicknesses are kept fixed (\tma$ = 10 $ nm and \tgma$= 30 $nm), while the spacer thickness is varied ($t_{\rm{spacer}} = 1,2,3,4,5 $ nm). 

The heterostructures used in our experiments are fabricated by low temperature molecular beam epitaxy on semi-insulating (001) GaAs substrates, after first depositing a 170 nm thick high-temperature grown GaAs buffer layer at 580$^{\circ}$C. We then grow a \gma~layer at a substrate temperature in the range 235$^{\circ}$C - 250$^{\circ}$C; the optimal substrate temperature depends on the Mn concentration. In particular, we note the use of distinct conditions during the growth of the highest Mn composition bilayer sample used in series B.\cite{Mack:2008tz} For trilayer samples, this is followed by the growth of p-doped GaAs:Be at the same substrate temperature.  The carrier concentration is approximately $3 \times 10^{19} {\rm{cm}}^{-3}$ as determined by Hall effect measurements of control samples. After the growth of the \gma~layer (or \gma/p-GaAs heterostructure), we lower the substrate temperature to $\sim$200$^{\circ}$C with the As shutter open to initiate the growth of a few monolayers of MnAs under As-rich conditions. We then raise the substrate temperature to $\sim$230$^{\circ}$C and continue growing MnAs; this procedure consistently yields MnAs in the ``type A'' orientation with the c-axis aligned with the $[\bar{1} 1 0]$ axis of the \gma~layer.

Cross-sectional transmission electron microscopy (TEM) shows an atomically abrupt and smooth interface between MnAs and \gma, despite a large lattice mismatch (Fig. 1(a)). Atomic force microscopy measurements of the top MnAs surface (Fig.1 (b)) shows that the freshly grown MnAs surface exhibits a relatively smooth surface with an rms roughness of about 1 nm. Some trenches with depth of $\sim$2 nm are observed as a result of the transition from 3D island growth to 2D layer growth. Note that the MnAs layer oxidizes quite readily, thus necessitating sample storage in vacuum for observing consistent physical properties with aging. The magnetic properties of the samples are characterized using a DC SQUID magnetometer. For temperature dependent measurements of the magnetization $M(T)$, samples are first cooled down from room temperature in a 20 kOe field applied along the easy axis of MnAs; unless otherwise stated, measurements are then taken while warming up in a field of 30 Oe. For magnetization hysteresis measurements, we focus here only on minor loops of the \gma~layers in order to determine the exchange field; the data are measured after first saturating the MnAs layer in a negative 20 kOe field.  All hysteresis loops unless noted are measured at 4.2 K.

\section{Exchange Spring Model of Bilayers}
We begin by discussing the interfacial exchange coupling in MnAs/\gma~bilayers. To calculate the exchange field experienced by the \gma~layer, we use a partial domain wall (PDW) model analogous to the one used in AFM/FM systems \cite{Mauri:1987:JAP} and hard/soft metallic FM bilayers.\cite{Guo:2002:APL} The magnetization of MnAs is considered to be fixed along its easy axis in the positive field direction along the [110] axis of \gma; this corresponds to the $[11 \bar{2} 0]$ direction of the hexagonal MnAs crystal. The magnetization of the \gma~layer is free to switch in an external magnetic field and we designate its direction with reference to the fixed MnAs magnetization, as illustrated in Fig. 1(c). We assume that a PDW of thickness $t_{1}$ is formed in the \gma~ layer near the interface. The angle between the MnAs magnetization and the \gma~ magnetization at the interface is defined as $\varphi_1$, while $\varphi_2$ is the angle between the MnAs magnetization and the bulk \gma~ magnetization.  Due to the strong coupling at the interface and the relatively strong anisotropy constant of MnAs, the interfacial spin alignment of the \gma~ layer should be very close to that in the MnAs layer.  For this strong interfacial coupling, where $\varphi_{1}\approx 0$  and $t_1 << t_{2}\approx t_{\rm{GMA}}$, the energy density per unit area can be written as:

\begin{eqnarray}
 E = 2\sqrt{AK}(1- cos\varphi_{2})-A_{ex}+K_{u}t_{\rm{GMA}}\sin^{2}\varphi_{2}\nonumber \\
   +1/4 K_{c}t_{\rm{GMA}} \cos^{2}2\varphi_{2}-HMt_{\rm{GMA}}\cos\varphi_{2}
\end{eqnarray}

The first term is the energy of the PDW, where $A$ is the spin stiffness of \gma, and $K$ is the effective anisotropy constant; the second term ($A_{ex}$) is the exchange coupling at the interface; the third and fourth terms are the uniaxial and biaxial anisotropy energy in \gma.  The terms $K_u$ and $K_c$ are the uniaxial and biaxial anisotropy constants, respectively.   The last term is the Zeeman energy where $H$ is the externally applied magnetic field and $M$ is the saturated magnetization of the \gma~ layer.  
When considering a strong cubic anisotropy, the energy minimum occurs at $45^{\circ}$ and $135^{\circ}$. Thus, we determine the two switching fields by using the following two conditions: $\frac {\partial^{2}E}{\partial \varphi^{2}_{2}}(\varphi_{2}=\pi/4) >0 ,  \quad
  \frac {\partial^{2}E}{\partial \varphi^{2}_{2}}(\varphi_{2}=3\pi/4) >0$.
This yields the following switching fields:
\begin{eqnarray}
   H > H_{C1} & = & \frac{-2\sqrt{2AK}-4K_{c}t}{\sqrt{2}Mt} \\
   H > H_{C2} & = & \frac{-2\sqrt{2AK}+4K_{c}t}{\sqrt{2}Mt}
\end{eqnarray}

The exchange field, given by $H_E = (H_{C1}+H_{C2})/2=-2\sqrt{AK}/Mt$, shows an inverse dependence on both the thickness and magnetization of the \gma~layer. In addition, the model also predicts that the coercive field, given by $H_c = (H_{C2}-H_{C1})=8K_c/\sqrt{2}M$, shows an inverse dependence on the magnetization of the \gma~layer.  We now test the validity of this model by studying the exchange coupling in MnAs/\gma~bilayers as a function of sample geometry and composition.

\section{Variation of exchange field and coercive field with \gma~ thickness in bilayers}
We first address the effect of varying the \gma~layer thickness (series A). Figure 2 (a) shows the temperature dependent remanent magnetization $M(T)$ in three bilayer samples with $t_{\rm{GMA}}=$ 30, 50, 80 nm, measured in a field of 30 Oe after cooling down from room temperature in a 20 kOe  field. We clearly observe two distinct FM phase transitions at $T_C \sim 75$ K for \gma~and $T_C \sim 318$ K for MnAs. The major magnetization hysteresis loops (data not shown) are similar to the data shown on other samples in a previous report,\cite{Zhu:2007kt} revealing two different coercivities for \gma~($\sim$100 Oe) and MnAs ($\sim$2 kOe). Based upon our SQUID measurements of these major hysteresis loops, we do not find any obvious indications of a biquadratic coupling. Figure 2 (b) shows the minor loops for two bilayers with different \gma~thicknesses; the displacement of the center of the minor loop is always opposite to the magnetization of the MnAs layer, indicating a ``negative exchange bias'' due to FM coupling between the two layers, where a parallel alignment of the two layers is favored. We find that the exchange field \he $ \sim (t_{\rm{GMA}})^{-1}$, in agreement with our model (Fig. 2(c)) and consistent with earlier measurements on a different sample series.\cite{Zhu:2007kt} With typical parameters of a 30 nm \gma~sample ($A\sim$0.4 pJ/m, $K\sim$0.3 kJ/m$^{3}$, $M\sim$16 emu/cm$^{3}$),\cite{Sugawara:2008:PRL} we calculate $H_E \sim 440$ Oe, which is reasonably close to the experimental value of 455 Oe. Our model also predicts that the coercive field \hc~ should be independent of \tgma; our data are in agreement with this, with \hc $\sim 100$ Oe for all three samples in Fig. 2(c).

Studies of ferromagnets exchange biased by an antiferromagnet typically show a correlation between \he~ and the coercive field \hc. This relationship can be studied by examining the temperature dependence of \he~ and \hc. For this purpose, we chose a MnAs/\gma~bilayer with high Mn concentration ($x= 0.16$) since the increased \tc ($= 160 \pm 5 $K) allows us to cover a wider temperature range. In contrast to the case of exchange biasing using an antferromagnet, we find that \hc~ and \he~ are not correlated: while \hc~ decreases with increasing temperature, \he~stays relatively constant (Fig. 2(d)).  The decrease in coercivity is expected and is also seen in non-exchange biased \gma~samples where it is attributed to a weakening of the magnetization and anisotropies of the \gma~layer.  However, the constant exchange field is surprising: our model predicts that as the magnetization of the \gma~layer decreases with increasing temperature, \he~ should increase.  However, we note that the anisotropy constants of \gma~ are also a function of temperature, decreasing with increasing temperature. Thus, a possible reason for our observation is that temperature dependence of both the magnetization and anisotropy cancel out any temperature variation in \he.
 
\section{Variation of exchange field and coercive field with \gma~ magnetization in bilayers}

A second prediction of the model is that \he~ should decrease inversely with the saturated magnetization ($M_{\rm{sat}}$) of the \gma~layer.  We test this prediction with a series of samples with varied Mn content (Series B).  Figure 3 (a) shows the minor loops of three samples with varying saturation of \gma~layer with nominal values of $x \approx 0.05, 0.07, 0.16.$ Figure 3(b) shows that the data are qualitatively consistent with the model (i.e. \he~ decreases with increasing $M_{\rm{sat}}$), but deviate from the predicted inverse dependence.  A plausible explanation for this deviation is that the changing Mn composition will also change the anisotropy constants, and not just saturated magnetization.  Also shown in Fig. 3(b) is the coercivity as a function of magnetization, which is qualitatively consistent with the ${(M_{\rm{sat}})}^{-1}$ dependence predicted by the model.

\section{Variation of exchange field with $\rm{MnAs}$ thickness in bilayers}
In the PDW model, we treat the MnAs layer as being essentially infinitely thick. In order to investigate the limitations of this assumption, we now address the behavior of bilayer samples in which we vary the thickness of the MnAs layer (\tma), keeping \tgma~fixed (series C).  We grew these samples by stopping the rotation of the wafer during the growth of the MnAs layer, allowing for a spatial variation of MnAs thickness across a single wafer.  We grew two wafers and cut each into 5 separate samples.  Figure 3(c) shows $M(T)$ for the first set of samples with 1.4 nm $\lesssim t_{\rm{MA}}  \lesssim 2$ nm, where we estimate \tma~using the saturated magnetization.  The second set has estimated values 3 nm $ \lesssim t_{\rm{MA}}  \lesssim 4$ nm.  Note that all these thicknesses are significantly thinner than our other sets of samples, which had $t_{\rm{MA}} \geq 8$ nm.  Figure 3(d) shows \he~vs. \tma, indicating that that the exchange field shows little dependence on \tma~for bilayer samples with at least 3 nm of MnAs, which is consistent with our model. However, Fig. 3(d) also shows that the exchange field rapidly decreases for very thin layers of MnAs (\tma $\lesssim 2$ nm).

The observed variation of the exchange field on the thickness of the biasing MnAs layer is reminiscent of the behavior in the conventional exchange biasing effect provided by an antiferromagnet.\cite{Nogues:1999:JMMM} Using the simple Meiklejohn-Bean model, exchange basing is obtained under the condition $K_{AFM} t_{AFM} >> A_{ex}$, where $K_{AFM}$ and  $t_{AFM}$ are the anisotropy and the thickness of the AFM layer and $A_{ex}$ is the interfacial exchange coupling. This model (and its more sophisticated extensions) thus predict that a critical thickness of the AFM layer is needed for exchange biasing with a value proportional to the ratio $\frac{A_{ex}}{K_{AFM}}$. Studies of AFM/FM bilayers have confirmed in some detail the expectations of this picture, showing both the quenching of exchange bias below a critical value as well as a saturation of the exchange bias at large AFM layer thickness.\cite{Lund_PhysRevB_2002} It is tempting to state that a similar picture could explain the variation of exchange field with the MnAs layer thickness: the only difference between our hard/soft FM bilayers and conventional AFM/FM bilayers is that the term describing the energy of the biasing layer depends upon the anisotropy of a ferromagnet rather than an antiferromagnet. Thus, our observation of an exchange field that saturates for rather small values of the biasing layer thickness (\tma~$\gtrsim 2$ nm) could be viewed as being qualitatively consistent with the relatively large anisotropy of the MnAs layer compared with the interfacial exchange coupling energy. 

\section{Interlayer exchange coupling in \gma/$\rm{GaAs/MnAs}$ trilayers: variation with thickness and doping of spacer layer}

Next, we address the propagation of the exchange coupling through a non-magnetic spacer by studying the behavior of MnAs/p-GaAs/\gma~trilayers (series D).  We have evidence that the exchange coupling between the two FM layers is mediated by holes in the spacer: Fig. 4(a) shows a comparison between minor hysteresis loops for two samples with a 3 nm spacer, one of which is {\it undoped} and the other doped with a nominal hole concentration of $3 \times 10^{19} {\rm{cm}}^{-3}$. We find no evidence for exchange biasing in the sample with the undoped spacer while the doped sample shows a clear shift, thus strongly suggesting that the exchange between the two magnetic layers is hole mediated. A systematic study of this coupling as a function of the doping density is beyond the scope of this paper. Instead, we focus on the spacer thickness (\tspacer) for a fixed p-doping level in the spacer. Figure 4 (b) shows both \he~and \hc~as a function of \tspacer, indicating that the exchange coupling becomes negligible when the spacer is larger than 5 nm (around the anticipated spin diffusion length in Be-doped GaAs). Note that we do not find any evidence for AFM exchange coupling over the entire space of parameters examined. The exchange field decays exponentially with the spacer thickness (as indicated by the fit in Fig. 4(b)), while the coercivity appears to have no spacer thickness dependence. The robustness of this behavior has been confirmed in a second set of trilayer samples (not shown).

We now discuss the monotonic ferromagnetic decay of the interlayer exchange within the context of the corresponding phenomenon in metallic multilayers where oscillatory interlayer exchange coupling is commonly observed in well prepared samples. In that case, the oscillatory dependence of the coupling on spacer thickness is understood using a model that relates the interlayer exchange to spin dependent reflection at interfaces and resultant quantum confined states within the spacer.\cite{Stiles:1999fm} As the spacer layer thickness is changed, the energy of these quantum well states changes. The oscillation period is then determined by the filling and emptying of these states as they pass through the Fermi energy of the spacer. The oscillation period is thus directly related to critical spanning vectors of the spacer layer Fermi surface and in a simple free electron gas picture is given by $\frac{\pi}{k_F}$. Such a model also predicts that the amplitude of the oscillatory coupling will be damped with an inverse dependence on \tspacer. If we apply these concepts to an ideal, disorder-free \gma/GaAs/MnAs trilayer, it is apparent that the oscillation period will be much longer than in a metallic system, simply because of the smaller carrier density in the semiconductor spacer.  For instance, for a hole density $p \sim 10^{19}$ cm$^{-3}$, the Fermi wave vector $k_F \sim 0.67$ nm$^{-1}$, so that the oscillation period is $\sim 5$ nm. Thus, even in an ideal sample, we would not expect to observe an oscillatory coupling over the spacer thicknesses studied in our experiments. It is however difficult to ignore the presence of disorder in our samples: the low temperature growth of the p-GaAs spacer results in a low carrier mobility and a short Drude mean free path ($\sim 4$ nm). Under these circumstances, the smearing of the Fermi surface can rapidly quench the oscillatory RKKY interaction. This could also account for the exponential decay in the amplitude of the coupling rather than the weaker inverse dependence on \tspacer~ expected in the ideal case.

Non-oscillatory FM coupling in multilayers can also arise from extrinsic effects: the most trivial example is that of direct FM coupling through pinholes.  This is ruled out by detailed TEM studies of our samples that show that the spacer layer is continuous with no obvious pinholes. Another possible extrinsic effect arises from the interdiffusion of magnetic ions into the nominally non-magnetic spacer. Experimental studies of Fe/Si/Fe trilayers with a thin ($< 1.6$ nm) undoped Si spacer showed exponentially decaying {\it antiferromagnetic} coupling with both bilinear and biquadratic terms.\cite{Strijkers_PhysRevLett.84.1812} This AF coupling was interpreted using a model that attributes the coupling to the polarization of paramagnetic loose spins in the spacer layer.\cite{SLONCZEWSKI:1993yk} The FM nature of the coupling observed in our samples is however contrary to the predictions of this model. Further, previous studies\cite{Wilson:2008wx} of interdiffusion in GaAs/MnAs superlattices suggest that the interdiffusion is limited to a several monolayers and is thus not extensive enough to produce the observed effect. Finally, an exponentially decaying exchange coupling was observed in exchange biased trilayer systems wherein an AFM biasing layer is separated from a FM layer by a noble metallic spacer layer.\cite{Gokemeijer_PhysRevLett.79.4270} Again, due to the vast differences in electronic structure and the density of states, it seems unlikely that there would be a common underlying physical mechanism that can describe the exponentially decaying coupling in both our samples and these metallic AFM/noble metal/FM trilayers.  

\section{Modification of \tc~ in \gma/$\rm{MnAs}$ heterostructures}

Finally, we address an intriguing possibility: is it possible that the exchange coupling between MnAs and \gma~could ``bootstrap'' the onset of ferromagnetism in the latter via a ``proximity'' effect. Recent x-ray magnetic circular dichroism studies have suggested that such a proximity effect results in room temperature ferromagnetism in a very thin region of \gma~within Fe/\gma~bilayers, although no direct evidence for such an effect is observed in magnetometry. \cite{Maccherozzi:2008kf} We have noticed that the growth of MnAs on top of \gma~consistently enhances \tc~of the latter typically by $\sim 25$ K compared to single epilayers of \gma~grown under similar conditions (an example is shown in Fig. 4(c)). The as-grown bilayer samples can show \tc~of up to 150 K, much higher than can normally be achieved before annealing. To better understand the nature of this effect and to see if it is intrinsic to the exchange coupling, we measured such bilayer samples after removing the top MnAs with a chemical etch.  Figure 4(d) shows the results of this control experiment: the \tc~of the \gma~layer remains elevated after etching the sample, instead of dropping as would be expected if the enhancement originated in a proximity effect. Our results thus suggest that the enhancement of \tc~is likely an extrinsic effect, stemming from very effective annealing of Mn interstitial defects during the overgrowth of MnAs.  To further confirm our hypothesis, we annealed the single epilayer control sample under similar conditions and found a very similar increase in \tc~(Fig. 4(d)).  We note that our magnetometry measurements cannot of course rule out the existence of exchange-enhanced ferromagnetism in a thin interfacial region of MnAs/\gma~bilayers.

\section{Summary}

In summary, we have reported a comprehensive study of exchange coupling in hybrid FM metal/semiconductor heterostructures. The interfacial exchange coupling between MnAs and \gma~results in a partial exchange spring configuration in the soft \gma~layer. We also show that this exchange coupling can propagate through a p-doped non-magnetic spacer layer, resulting in an interlayer exchange coupling in trilayer configurations. Using a metallic FM layer to exchange bias \gma~offers a new testbed for studying exchange coupling between FM metals and semiconductors, and it also provides a model system to study spin-dependent transport in non-uniform magnetization configurations.\cite{Foros:2008gv,Levy:1997lr}
 As an engineering tool, it opens up opportunities for tailoring the coercivity of FM semiconductors for proof-of-concept device applications.

This research is supported by the ONR MURI program N0014-06-1-0428. This work was performed in part at the Penn State Nanofabrication Facility, a member of the NSF National Nanofabrication Infrastructure Network. We thank Michael Flatt\'e and Mark Stiles for insightful discussions.

\newpage
\begin{figure}[]
\begin{center}
\includegraphics[]{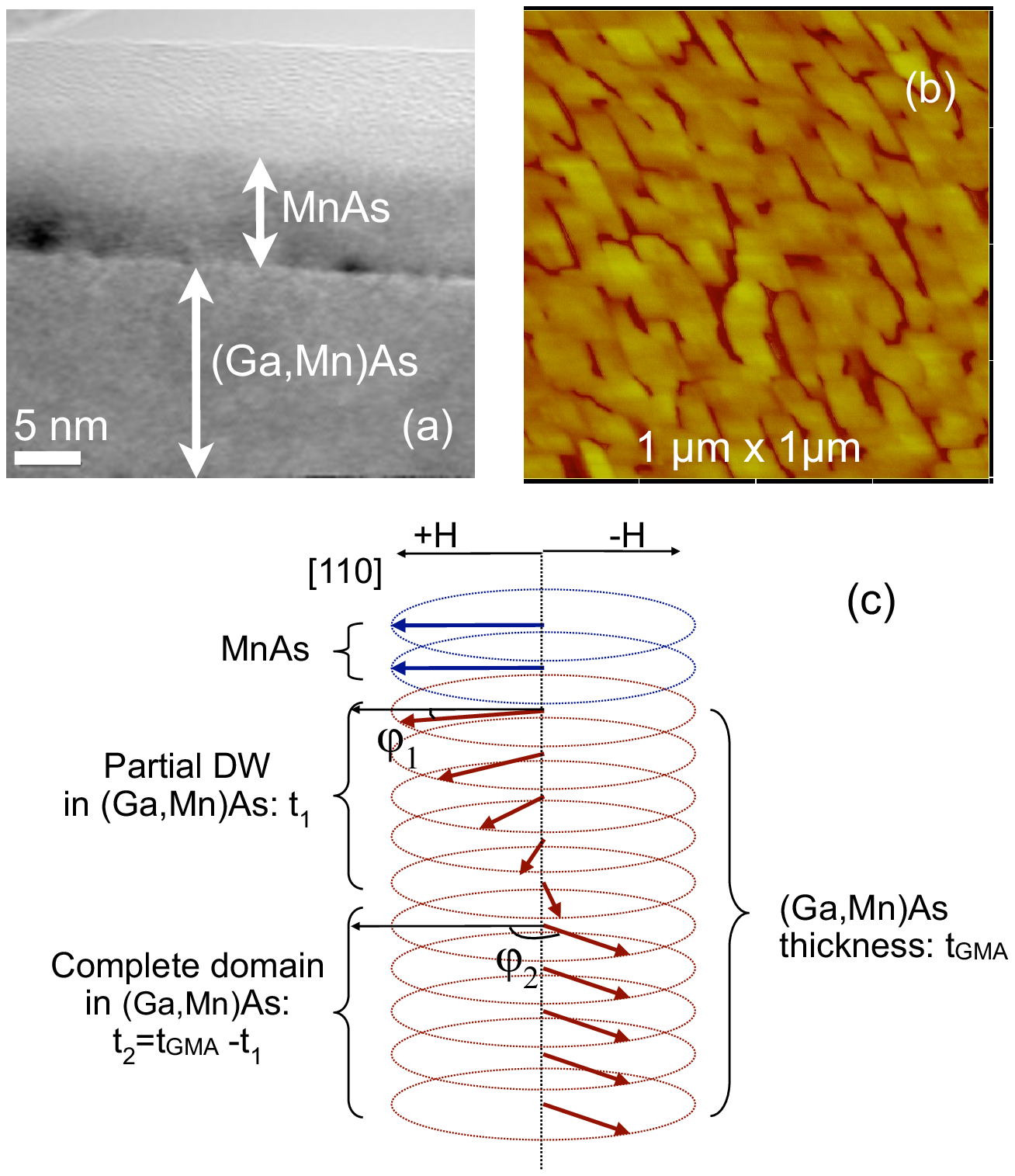}
\caption{(a) HRTEM images of a bilayer sample. (b) Atomic force microscope image of the top MnAs layer. (c) Depiction of a partial domain wall configuration in \gma, with spins continuously rotating as a function of the distance from the interface. Beyond a certain depth $t_{1}$, a complete domain ($t_{2}$) forms.}
\label{fig:bilayer_TEM_AFM}
\end{center}
\end{figure}

\newpage
\begin{figure}[]
\begin{center}
\includegraphics[]{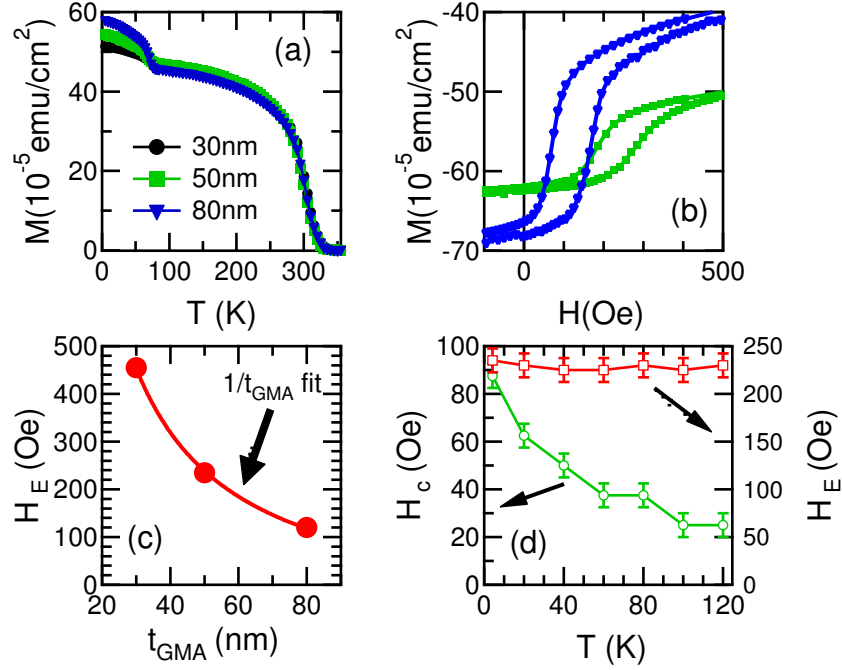}
\caption{(a) Temperature-dependent magnetization $M(T)$ for bilayer samples with three different \gma~thicknesses. (b) Minor hysteresis loop for the bilayer samples with \tgma = 50 nm (green squares) and 80 nm (blue triangles). Data are taken at $T = 4.2$ K. (c) Exchange field versus \gma~ thickness, showing that \he~$\propto$~(\tgma$)^{-1}$. Data are taken at $T = 4.2$ K. (d) Coercivity and exchange field as a function of temperature for a high Mn (x=16) composition MnAs/\gma~bilayer with \tgma = 50 nm.}
\label{fig:Magnetization_bilayer}
\end{center}
\end{figure}

 \newpage
\begin{figure}[]
\begin{center}
\includegraphics[]{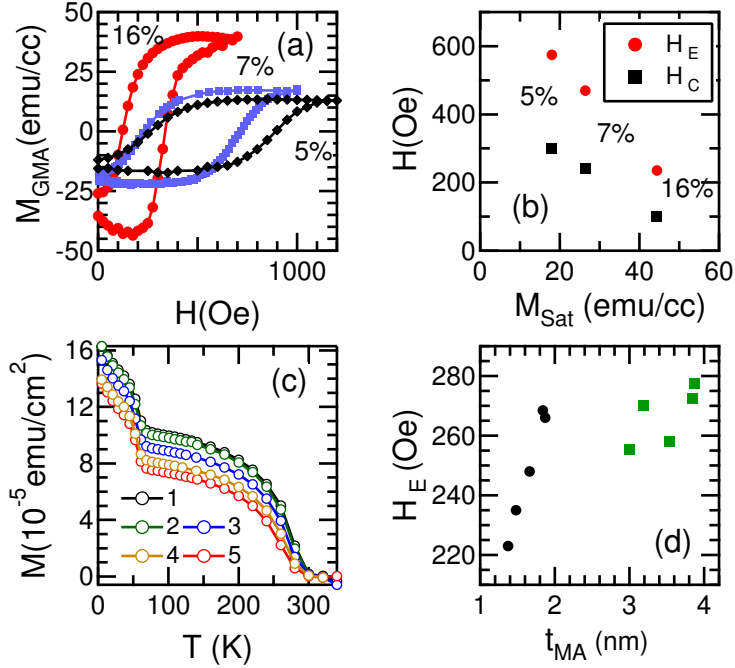}
\caption{(a) Minor hysteresis loops of three bilayer samples with varying Mn concentration.  The magnetization is shown per unit volume with the MnAs signal subtracted out. Data are taken at $T = 4.2$ K.  (b) Exchange field and coercivity as a function of saturated magnetization. Data are taken at $T = 4.2$ K.  (c) Temperature-dependent magnetization $M(T)$ for bilayer samples with different MnAs layer thicknesses. The magnetization is shown per unit area. (d) Exchange field versus MnAs layer thickness showing critical thickness between 2 and 3 nm. Data are taken at $T = 4.2$ K. }
\label{fig:PDW_model}
\end{center}
\end{figure}

\newpage
\begin{figure}[]
\begin{center}
\includegraphics[]{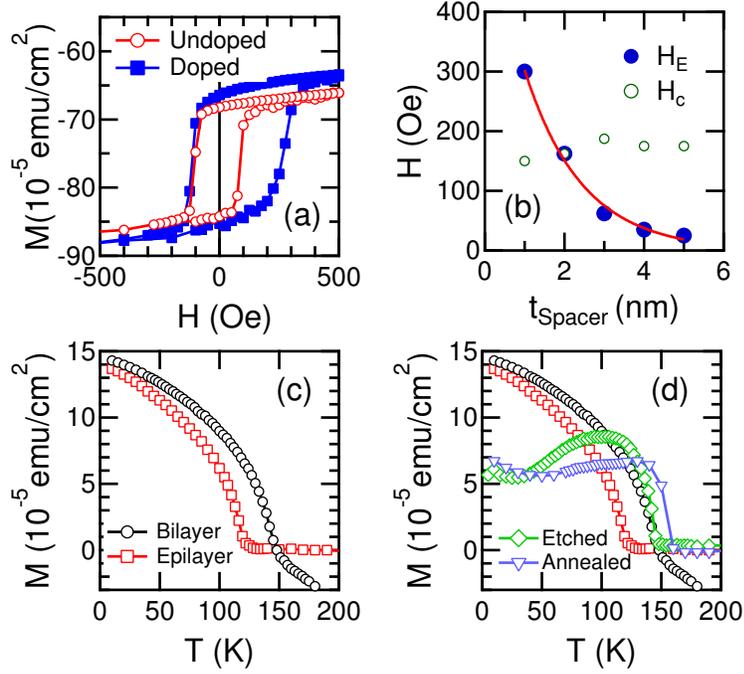}
\caption{(a) Minor loops for two 3 nm trilayer samples, one with an undoped spacer and one with a Be doped spacer. Data are taken at $T = 4.2$ K. (b) Exchange field (\he) and coercivity (\hc) in \gma/p-GaAs/MnAs trilayers versus p-GaAs spacer layer thickness. Data are taken at $T = 4.2$ K.  The solid line shows a fit of the variation of \he~ vs. \tspacer to an exponential decay: \he~$\propto \exp (-\alpha t_{\rm{spacer}})$. (c) Temperature-dependent magnetization $M(T)$ for bilayer and single layer control sample. These measurements are taken while warming up in a field of 200 Oe.  (d) Same as in (c) with the added etched sample showing that \tc~did not change and with added annealed control sample showing very similar increase of \tc. The unusual shape of the $M(T)$ at lower temperatures results from temperature dependent changes in the easy axis.}
\label{fig:Magnetization_t}
\end{center}
\end{figure}




\end{document}